# Mercury's Protoplanetary Mass

J. Marvin Herndon

Transdyne Corporation
San Diego, California 92131 USA

September 30, 2004

## Abstract

Major element fractionation among chondrites has been discussed for decades as ratios relative to Si or Mg. Recently, by expressing ratios relative to Fe, I discovered a new relationship admitting the possibility that ordinary chondrite meteorites are derived from two components, a relatively oxidized and undifferentiated, *primitive* component and a somewhat differentiated, *planetary* component, with oxidation state like the highly reduced enstatite chondrites, which I suggested was identical to Mercury's complement of "lost elements". Here, on the basis of that relationship, I derive expressions, as a function of the mass of planet Mercury and the mass of its core, to estimate the mass of Mercury's "lost elements", the mass of Mercury's "alloy and rock" protoplanetary core, and the mass of Mercury's gaseous protoplanet. Although Mercury's mass is well known, its core mass is not, being widely believed to be in the range of 70-80 percent of the planet mass. For a core mass of 75 percent, the mass of Mercury's "lost elements" is about 1.32 times the mass of Mercury, the mass of the "alloy and rock" protoplanetary core is about 2.32 times the mass of Mercury, and the mass of the gaseous protoplanet of Mercury is about 700 times the mass of Mercury. Circumstantial evidence is presented in support of the supposition that Mercury's complement of "lost elements" is identical to the *planetary* component of ordinary-chondrite-formation.

**Introduction**

As first noted by Urey (1951), the planet Mercury consists mainly of iron. Sometime during its formation, a significant portion of Mercury's original complement of elements was lost; volatilization of silicates during the Sun's formation has been suggested (Bullen 1952; Cameron 1985; Ringwood 1966; Urey 1952). Recently, for the first time I addressed the ultimate fate of Mercury's lost elements through implications derived from a new fundamental relationship among the major elements (Fe, Mg, Si) of chondritic meteorites (Herndon 2004b). Here I use that relationship to derive new implications on the mass of Mercury's "lost elements" component and the original mass of the protoplanet from which Mercury formed.

For a time during the 1940s and 1950s, gaseous protoplanets were discussed (Kuiper 1951a; Kuiper 1951b; Urey 1952), but the idea of protoplanets was largely abandoned in favor of the so-called "standard model" of solar system formation, based upon the concept that grains condensed from diffuse nebula gases at a pressure of about $10^{-5}$ bar, and were then agglomerated into successively larger pebbles, rocks, planetesimals and, ultimately, planets (Wetherill 1980).

Recently, I showed that the "standard model" of solar system formation is *wrong* because it would yield terrestrial planets having insufficiently massive cores, a profound contradiction to what is observed (Herndon 2004d). The "standard model" of solar system formation is *wrong* because the popular underlying "equilibrium condensation" model is *wrong*, the Earth in its composition is *not* like an ordinary chondrite meteorite as frequently assumed, and condensates from nebula gases at $10^{-5}$ bar would be *too* oxidized to yield planetary cores of sufficient mass. Instead, within the framework of present knowledge, the concept of planets having formed by raining out from the central regions of hot, gaseous protoplanets, as revealed by Eucken (1944), appears to be consistent with the observational evidence (Herndon 2004c; Herndon 2004d). The nature of that condensate appears to have the same composition and state of oxidation as an enstatite chondrite meteorite.

I have shown from fundamental ratios of mass (Table 1) that the Earth as a whole, and especially, the endo-Earth, the inner 82% comprising the lower mantle and core, has the composition and oxidation state like a highly reduced enstatite chondrite meteorite (Herndon 1993; Herndon 1996; Herndon 2004a). As inferred from reflectance spectroscopy (Vilas 1985), the near absence of oxidized iron in the regolith of Mercury indicates that it too has a similar, highly reduced state of oxidation.

Major element fractionation among chondrites has been discussed for decades as ratios relative to Si or Mg. Expressing ratios relative to Fe led me to a new fundamental relationship admitting the possibility that ordinary chondrite meteorites are derived from two components: one is a relatively undifferentiated, *primitive* component, oxidized like the CI or C1 chondrites; the other is a somewhat differentiated, *planetary* component, with oxidation state like the highly reduced enstatite chondrites (Herndon 2004b). Because of the state of oxidation of the *planetary* component, the ubiquitous deficiencies of refractory siderophile elements in ordinary chondrites which are reflected in the *planetary* component, and with due consideration for the relative masses involved, I suggested that the *planetary* component of ordinary-chondrite-formation consists of planet Mercury's complement of "lost elements". In the following I derive estimates of the mass of Mercury's "lost elements" and the mass of the Mercury protoplanet, assuming that the *planetary* component of ordinary-chondrite-formation is in fact identical to Mercury's "lost elements".

## Theoretical Considerations

Neglecting for the moment the great quantities of protoplanetary H, He, and other volatile elements and compounds, the mass of the readily condensable "rock and alloy" component of the mass of the protoplanet of Mercury, $M_{Pp}$, is given by

$$M_{Pp} = M_{Mc} + M_{Mm} + M_{Lc} + M_{Lm} \qquad (1)$$

where $M_{Mc}$ is the mass of planet Mercury's core, $M_{Mm}$ is the mass of Mercury's mantle, $M_{Lc}$ is the mass of the core-forming component of Mercury's lost elements, and $M_{Lm}$ is the mass of the mantle-forming component of Mercury's lost elements.

From fundamental relationships, the mass ratio of the lower mantle to core of the Earth is virtually identical to the ratio of silicate to alloy of the Abee enstatite chondrite, a value of 1.4272. So, it follows, that the corresponding components of the readily condensable "rock and alloy" portion of the mass of the protoplanet of Mercury should be likewise related, thus

$$\frac{(M_{Mm} + M_{Lm})}{(M_{Mc} + M_{Lc})} = 1.4272 \qquad (2)$$

I have suggested that the *planetary* component of ordinary-chondrite-formation is identical to the "lost elements" portion of the readily condensable "rock and alloy" component of the mass of the protoplanet of Mercury. From Herndon (2004b), the molar Mg/Fe ratio of the planetary component is 3.0956, which corresponds to a Mg/Fe mass ratio equal to 1.3472.

Note that in enstatite-chondrite-matter Mg occurs mainly in the silicates, while Fe occurs almost exclusively in the alloy portion corresponding to the core. Thus, the mass ratio of Mg to the sum of the silicate-forming-elements in enstatite-chondrite-matter, from the Abee enstatite chondrite, is like that of the corresponding ratio of the silicate-forming portion of Mercury's "lost elements" component:

$$\frac{Mg}{M_{Lm}} = 0.1877$$

Likewise, the mass ratio of Fe to the sum of the alloy-forming-elements is

$$\frac{Fe}{M_{Lc}} = 0.7579$$

Combining and recalling that for the *planetary* component Mg/Fe = 1.3472,

$$1.3472 = \frac{Mg}{Fe} = \frac{0.1877\ M_{Lm}}{0.7579\ M_{Lc}}$$

Thus,

$$M_{Lm} = 5.4398\ M_{Lc}$$

Substitution into (2) gives

$$M_{Lc} = 0.3557\ M_{Mc} - 0.2492\ M_{Mn}$$

So that,

$$M_{Lm} = 1.9349\ M_{Mc} - 1.3556\ M_{Mn}$$

Thus, from (1)

$$M_{Pp} = M_{Mc} + M_{Mm} + 0.3557\ M_{Mc} - 0.2492\ M_{Mn} + 1.9349\ M_{Mc} - 1.3556\ M_{Mn}$$

$$M_{Pp} = 3.2906\ M_{Mc} - 0.6048\ M_{Mn} \quad (3)$$

## Results and Discussion

The results of computations made using the above equations are shown in Fig. 1. Only the mass of planet Mercury and the mass of its core would be sufficient for a unique determination of the mass of the "rock and alloy" portion of its protoplanet and Mercury's "lost elements" component. Although the mass of planet Mercury is well known, considerable uncertainty presently exists as to the precise mass of its core. There is widespread belief that Mercury's core accounts for 70-80 percent of the planet mass. The great uncertainty in core mass arises primarily from the uncertainty in Mercury's moment of inertia, which will be measured during NASA's Messenger mission.

The parameters presented in Fig. 1 are shown as well for certain specific core masses in Table 1, expressed relative to planet Mercury's mass of $3.3022 \times 10^{23}$ kg. In addition,

Table 1 contains estimates of the mass of the gaseous protoplanet from which Mercury formed, which from solar abundances is about 300 times the mass of the "rock and alloy" portion of that protoplanet.

The calculations made in this paper are based upon the supposition that Mercury's "lost elements" component is in fact one of the two components from which the ordinary chondrites were formed. The circumstantial evidence in support of that supposition follows: *i*) the *planetary* component of ordinary-chondrite-formation is partially differentiated, highly-reduced, enstatite-chondrite matter like Mercury, *ii*) the relative masses involved are consistent, *iii*) the ordinary chondrites formation location is thought to be mainly within the asteroid belt, *iv*) the primordial gases were stripped that were originally associated with the protoplanets of the terrestrial planets, *v*) Mercury, being closest to the Sun, would have experienced the most direct assault by the high temperatures and/or by the violent activity during some early super-luminous solar event, such as the T-Tauri phase solar wind associated with the thermonuclear ignition of the Sun, *vi*) the ordinary chondrites contain chondrules, small objects that appear like droplets from a fiery rain that obtained their shape from the surface tension of the melt while in space, and, *vii*) the ordinary chondrites and the *planetary* component are deficient in siderophile refractory elements, an indication that the *planetary* component came from a partially differentiated, heterogeneously accumulated body, consistent with Mercury's planetary core having begun to rain out from its protoplanet (Herndon 2004d). Taken together these form an internally consistent picture.

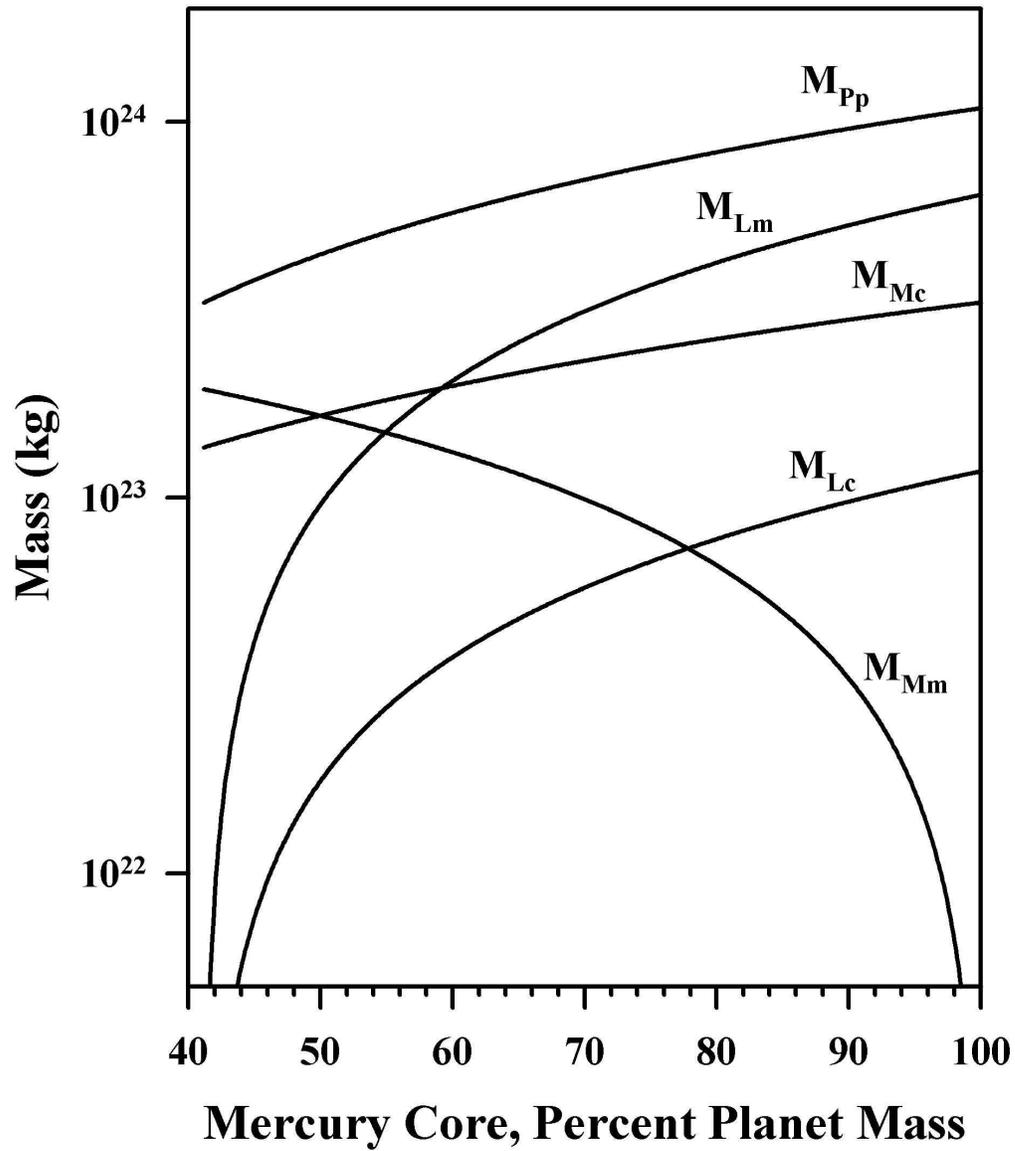

**Figure 1**. Lacking knowledge of the precise value for the mass of Mercury's core, the results of calculations are shown for a range of masses. Symbols for mass curves: $M_{Pp}$ "rock and alloy" protoplanet; $M_{Lm}$ silicate component of "lost elements"; $M_{Mc}$ Mercury core; $M_{Lc}$ alloy component of "lost elements"; $M_{Mm}$ Mercury mantle.

**Table 1**. Fundamental mass ratio comparison between the endo-Earth (core plus lower mantle) and the Abee enstatite chondrite (Herndon 2004a).

| Fundamental Earth Ratio | Earth Ratio Value | Abee Ratio Value |
| --- | --- | --- |
| lower mantle mass to total core mass | 1.49 | 1.43 |
| inner core mass to total core mass | 0.052 | *theoretical* <br> 0.052 if $Ni_3Si$ <br> 0.057 if $Ni_2Si$ |
| inner core mass to (lower mantle+core) mass | 0.021 | 0.021 |
| core mean atomic mass | 48 | 48 |
| core mean atomic number | 23 | 23 |

**Table 2**. Numerical values from Fig. 1 calculated for certain specific core masses are shown relative to Mercury's mass of $3.3022 \times 10^{23}$ kg.

| Mercury Core | Mercury Mantle | Lost Elements | Protoplanet Rock/Alloy | Protoplanet Gaseous |
|---|---|---|---|---|
| 0.45 | 0.65 | 0.148 | 1.15 | 344 |
| 0.50 | 0.50 | 0.343 | 1.34 | 403 |
| 0.55 | 0.45 | 0.538 | 1.54 | 461 |
| 0.60 | 0.40 | 0.732 | 1.73 | 520 |
| 0.65 | 0.35 | 0.927 | 1.93 | 578 |
| 0.70 | 0.30 | 1.122 | 2.12 | 637 |
| 0.75 | 0.25 | 1.317 | 2.32 | 695 |
| 0.80 | 0.20 | 1.512 | 2.51 | 754 |
| 0.85 | 0.15 | 1.706 | 2.71 | 812 |
| 0.90 | 0.10 | 1.901 | 2.90 | 869 |
| 0.95 | 0.05 | 2.096 | 3.10 | 929 |